\documentclass[a4paper,12pt, epsfig]{article}
\usepackage{epsfig}
\usepackage{amssymb}
\usepackage{amsfonts}
\usepackage{amsmath}
\usepackage{subfig}
\usepackage{wrapfig}

\newskip\humongous \humongous=0pt plus 1000pt minus 1000pt

\newif\ifdtup

\catcode`\@=11

\@addtoreset{equation}{section}
\def\theequation{\thesection.\arabic{equation}}

\def\@normalsize{\@setsize\normalsize{15pt}\xiipt\@xiipt
\abovedisplayskip 14pt plus3pt minus3pt%
\belowdisplayskip \abovedisplayskip
\abovedisplayshortskip \z@ plus3pt%
\belowdisplayshortskip 7pt plus3.5pt minus0pt}

\def\small{\@setsize\small{13.6pt}\xipt\@xipt
\abovedisplayskip 13pt plus3pt minus3pt%
\belowdisplayskip \abovedisplayskip
\abovedisplayshortskip \z@ plus3pt%
\belowdisplayshortskip 7pt plus3.5pt minus0pt
\def\@listi{\parsep 4.5pt plus 2pt minus 1pt
     \itemsep \parsep
     \topsep 9pt plus 3pt minus 3pt}}

\relax

\catcode`@=12

\catcode`\@=11

\def\section{\@startsection{section}{1}{\z@}{3.5ex plus 1ex minus
   .2ex}{2.3ex plus .2ex}{\large\bf}}

\def\thesection{\arabic{section}}
\def\thesubsection{\arabic{section}.\arabic{subsection}}

\def\appendix{\setcounter{section}{0}
 \def\thesection{Appendix \Alph{section}}
 \def\thesubsection{\Alph{section}.\arabic{subsection}}
 \def\theequation{\Alph{section}.\arabic{equation}}}

\def\SymBoxes#1#2#3#4{\newdimen\un@t \un@t#3%
\raisebox{#1}{\rule{#2\un@t}{#4}\hskip-#2\un@t
\@tempdimb\un@t \advance\@tempdimb by-#4\@tempcntb#2\relax%
\@whilenum{\@tempcntb>0}\do{
\rule{#4}{\un@t}\hskip\@tempdimb \advance\@tempcntb by\m@ne}%
\hskip-#2\un@t \rule[\un@t]{#2\un@t}{#4}%
\rule[\un@t]{#4}{#4}\hskip-#4
\rule{#4}{\un@t}}\hskip-#4}                

\begin{document}

\newcommand{\beq}{\begin{equation}}
\newcommand{\eeq}{\end{equation}}
\newcommand{\bea}{\begin{eqnarray}}
\newcommand{\eea}{\end{eqnarray}}
\newcommand{\beas}{\begin{eqnarray*}}
\newcommand{\eeas}{\end{eqnarray*}}
\newcommand{\defi}{\stackrel{\rm def}{=}}
\newcommand{\non}{\nonumber}
\newcommand{\bquo}{\begin{quote}}
\newcommand{\enqu}{\end{quote}}
\renewcommand{\(}{\begin{equation}}
\renewcommand{\)}{\end{equation}}
\def\IZ{{\mathbb Z}}
\def\IR{{\mathbb R}}
\def\IC{{\mathbb C}}
\def\IQ{{\mathbb Q}}
\def\IP{{\mathbb P}}

\def \eqn#1#2{\begin{equation}#2\label{#1}\end{equation}}
\def\de{\partial}
\def\Tr{ \hbox{\rm Tr}}
\def\H{ \hbox{\rm H}}
\def\HE{ \hbox{$\rm H^{even}$}}
\def\HO{ \hbox{$\rm H^{odd}$}}
\def\K{ \hbox{\rm K}}
\def\Im{ \hbox{\rm Im}}
\def\Ker{ \hbox{\rm Ker}}
\def\const{\hbox {\rm const.}}
\def\o{\over}
\def\im{\hbox{\rm Im}}
\def\re{\hbox{\rm Re}}
\def\bra{\langle}\def\ket{\rangle}
\def\Arg{\hbox {\rm Arg}}
\def\Re{\hbox {\rm Re}}
\def\Im{\hbox {\rm Im}}
\def\exo{\hbox {\rm exp}}
\def\diag{\hbox{\rm diag}}
\def\longvert{{\rule[-2mm]{0.1mm}{7mm}}\,}
\def\a{\alpha}
\def\dag{{}^{\dagger}}
\def\tq{{\widetilde q}}
\def\p{{}^{\prime}}
\def\W{W}
\def\N{{\cal N}}
\def\hsp{,\hspace{.7cm}}
\newcommand{\C}{\ensuremath{\mathbb C}}
\newcommand{\Z}{\ensuremath{\mathbb Z}}
\newcommand{\R}{\ensuremath{\mathbb R}}
\newcommand{\rp}{\ensuremath{\mathbb {RP}}}
\newcommand{\cp}{\ensuremath{\mathbb {CP}}}
\newcommand{\vac}{\ensuremath{|0\rangle}}
\newcommand{\vact}{\ensuremath{|00\rangle}}
\newcommand{\oc}{\ensuremath{\overline{c}}}
\begin{titlepage}
\begin{flushright}
ULB-TH/08-30\\
\end{flushright}
\bigskip
\def\thefootnote{\fnsymbol{footnote}}

\begin{center}
{\Large {\bf Baryon Dissociation in a Strongly Coupled Plasma }}
\end{center}

\bigskip
\begin{center}
{\large  Chethan
KRISHNAN\footnote{\texttt{Chethan.Krishnan@ulb.ac.be}} }
\end{center}

\renewcommand{\thefootnote}{\arabic{footnote}}

\begin{center}
\vspace{1em}
{\em  { International Solvay Institutes,\\
Physique Th\'eorique et Math\'ematique,\\
ULB C.P. 231, Universit\'e Libre
de Bruxelles, \\ B-1050, Bruxelles, Belgium\\}}

\end{center}

\noindent
\begin{center} {\bf Abstract} \end{center}
Using the dual string theory, we study a circular baryonic
configuration in a wind of strongly coupled ${\cal N}=4$ Yang-Mills plasma
blowing in the plane of the baryon, before and after a quark has
dissociated from
it. A simple enough model that captures many interesting features is
when there are four quarks in the baryon. As a step towards phenomenology,
we compare representative dissociated configurations, and make some comments about their energetics and other properties. Related results that we find include the observation that the screening length formula $L_s T \sim (1-v^2)^{1/4}$ obtained previously for other color singlet configurations, is robust for circular baryons as well.

\vspace{1.6 cm}

\vfill

\end{titlepage}
\bigskip

\hfill{}
\bigskip

\tableofcontents

\setcounter{footnote}{0}
\section{\bf Introduction and Conclusion} \label{intro}

\noindent
There is a possibility that screening of heavy quark baryons in a wind of strongly coupled plasma might be experimentally accessible at RHIC or LHC\footnote{We emphasize that so far heavy quark baryons have not been observed either in elementary collisions or heavy ion collisions.}. Unfortunately, this is a deeply
non-perturbative scenario in standard QCD, and therefore essentially
out of theoretical control. Interestingly enough, semi-quantitative
features of baryon screening
can be computed using a dual string theory through the AdS/CFT
correspondence \cite{adscft}. This has resulted in some non-trivial progress in the understanding of screening phenomena in strongly coupled plasmas \cite{Liu:2006nn, elena, Liu:2006he, A}.

In a wind of plasma, the screening is direction-dependent, and it stands to reason
that when the baryon dissociates, there will be preferred directions in
which this can happen first. The purpose of this paper is to explore this
possibility by computing some basic energetics. This should be taken as a small step towards phenomenology.

As a warmup, we first compute the (regulated) energy of a circular baryonic configuration with $N_c=4$ quarks\footnote{The choice 4 is obtained by optimizing between non-triviality and simplicity.} (attached in the bulk to a D5-brane baryon vertex wrapping the $S^5$), moving in a hot, strongly coupled, ${\cal N}=4$ plasma. For a generic baryon configuration, this can be done using worldsheet string theory in $AdS_5 \times S^5$, with a black hole in the interior. Our results about these baryons also serve as a test of robustness of the known results in the literature. After this, we calculate the energy of the configuration after one of the quarks has dissociated from the baryon and is well-separated. Both the quark, and what remains of the baryon, will have trailing strings reaching down to the black hole horizon. There are many such dissociated configurations, and we will try to get some intuition for them by considering some special cases. Among these will be two extreme scenarios: one when the remaining three quarks are spaced equidistantly in a line along the wind (case I), and the other when the quarks (again spaced equidistantly along a line) are perpendicular to the wind (case II). The ``dynamics" of the system is sufficiently stringent that for the configurations we consider, we will be able to explicitly do the computations without getting tangled up in too many details.

 We stress that our aim is not to identify preferred dissociation channels and do any detailed phenomenology. This would be a tall order. Our aim is merely to see whether something can be said about energetics of transverse configurations vs. longitudinal ones. The two specific cases we consider are chosen with this specific purpose in mind, and they are not supposed to necessarily be the preferred dissociation products. In fact, one might think that the ``natural" dissociated configurations are instead the ones we consider in Appendix C. But this is so only if one assumes that the remaining quarks stay on the circle. The problem here is that ``natural" is not very well-defined, because the dynamics is not under control and all we are dealing with are static configurations.

In any event, we find that the energy of the configuration with quarks parallel to the wind (case I) is at a higher energy than case II, and therefore it is tempting to speculate that for baryons in a wind of plasma, the quarks {\em along} the wind will dissociate first. We emphasize that the computations we do should not be taken as a proof of this claim, even though we believe they are suggestive. In particular, it is not clear what configurations constitute extreme cases. It is a complicated dynamical question whether a configuration would prefer to change the dimension (i.e., change the length $L$) and/or change the orientation in order to lower its energy. But it can still be instructive to have a comparison of identical configurations with different orientations: naively, one might think that for a given dimension, a transverse configuration is at a higher energy because it has more cross-section to the wind. Our results show that this is not the case. Also, there are speculations and comments in the literature about the relative energetics of transverse-vs.-longitudinal cases. Such statements make most sense only if one assumes that we are comparing two configurations of identical dimensions, and this is what we do.


In general, the features of a dissociated baryon are likely to depend strongly on the specific configurations under consideration, and making generic claims is difficult. To emphasize this, we compute the energetics of some other dissociated configurations in an appendix.

A more detailed computation where the number of quarks and the possible dissociation channels are increased will certainly be useful in shedding more light on similar questions. It would be interesting to do a full scan in the the space of allowed dissociated configurations, but this is numerically a more demanding problem than what we have undertaken in this paper. It should also be pointed out that conclusive evidence might require an understanding of the dynamics of dissociation patterns which our static calculation is blind to: in principle, it is always possible for instance that the configuration can rotate or dilate during or after a dissociation process.

In the course of the computations in this paper we learn a few things about dragging and non-dragging objects and these will be elaborated upon as and when they arise. Among these is the observation that the screening length formula found previously in the literature for other color singlet configurations, $L_s T \sim (1-v^2)^{1/4}$, is valid for circular baryons as well. This is interesting because the details of the configurations and the computations are quite different in our case. Another thing we find is that the plots of $E$ vs. $L$ for non-dragging objects exhibit a cusp while those of dragging configurations exhibit a loop, and we speculate that the area of the loop is a measure of the drag of the configuration. This could potentially be a measure of the energy loss.

In this paper, we only consider the simplest possible scenario: a baryon that is simulated by a circular configuration of four (external) quarks in the maximally symmetric gauge theory, with the
added simplification that the plasma is flowing in the plane of the baryon. But it is possible to consider generalizations of the results here to more generic baryon configurations, less special gauge theories and perhaps more generic wind directions. It would be interesting to see how generic the results are. Some papers that are relevant (from various angles) to hot strongly coupled QCD are \cite{Herzog,k-3,H2,F,k-2,k-1,k0,k1,k2,k3,k4,k5,k6}.

\section{Baryon Screening and the AdS Black Hole}

Baryons are made out of fundamental quarks in QCD, but the field content
of the ${\cal N}=4$  supersymmetric Yang-Mills theory consists of gluons,
gluinos and
scalars (all in the adjoint); but not quarks. So in order to
model QCD phenomena, we introduce baryons that are constructed out of
external quarks. If the
${\cal N}=4$ SYM theory has $N_c$ colors, the baryons will be constructed
from $N_c$ such external quarks.

The dynamics of baryons in
the gauge theory is captured in the dual $AdS_5\times
S^5$ string theory through the introduction of the so-called baryon vertex
\cite{Witten:1998xy}. The claim is that baryons in the gauge
theory are dual to configurations that involve a D5-brane wrapping an
$S^5$ in the bulk, with all the heavy external quarks in the
boundary baryon being linked to it through fundamental strings (all of
which are of the same orientation).

The way in which we make predictions for a baryon moving in the plasma is
by boosting to the rest frame of the baryon and letting the plasma move
instead. In the dual picture, we look for static baryon configurations in
the boosted bulk metric. In the course of this paper, we will be
exclusively working with the case of ${\cal N}=4 \ {\rm SYM}$, with the
plasma at a temperature $T$. Finite temperature implies that the
asymptotics of the bulk is still $AdS_5 \times S^5$ but in the
interior we have to change the metric to include a black hole whose
Hawking temperature is $T$ \cite{hawking-page}. Before the boost, this
bulk AdS black
hole metric takes the
form
\beq
 ds^2 =  - f(r)  dt^2 +  \frac{r^2}{ R^2} d\vec x^2 +  \frac{dr^2}{f(r)} +
R^2 d \Omega_5^2,
\label{ZeroVelocityMetric}
\eeq
with
\beq
f(r)=\frac{r^2}{R^2}\left(1-\frac{r_0^4}{r^4}\right).
\label{fdefn}
\eeq
The asymptotic boundary where the field theory lives is supposed to be at
$r\rightarrow \infty$, and is spanned by $\vec x =\{x_1, x_2, x_3\}$ and
the bulk-time $t$. In the above, $r_0$ is the black hole horizon and the
temperature $T$ is fixed by the Hawking relation $T=\frac{r_0}{\pi R^2}$.
The standard CFT to AdS correspondence is usually expressed by
taking the defining parameters of the gauge theory to be $\lambda \
(\equiv
g^2_{YM}N_c)$ and $N_c$. Then, the bulk data is related to the boundary
data through Maldacena's famous relations
\beq
\frac{\lambda}{N_c}=4\pi g_s, \  \sqrt{\lambda}=\frac{R^2}{\alpha'},
\eeq
with $g_s$ the string coupling and $\frac{1}{2\pi \alpha'}$ the
worldsheet tension. The Maldacena conjecture can be taken as the claim
that string propagation with these parameters in the AdS background (with
a background five-form flux controlled by $R$) is just another description
of the gauge theory.

We will take the plasma wind to be in the $x_3$-direction after the boost,
and the velocity and rapidity are related by $v=-\tanh \eta$. The boosted
metric takes the form
\begin{equation} \label{boostedmetric}
ds^2=-A dt^2 +
2B\,dt\,dx_3+
C\,dx^2_3+\frac{r^2}{R^2}\left(dx^2_1+dx^2_2\right)+\frac{1}{f(r)}dr^2+R^2
d\Omega^2_5\,.
\end{equation}
The various quantities are fixed by,
\begin{equation} \label{eq:2}
A=\frac{r^2}{R^2}\left(1-\frac{r^4_1}{r^4}\right), \qquad
B=\frac{r^2_1 r^2_2}{r^2 R^2}, \qquad
C=\frac{r^2}{R^2}\left(1+\frac{r^4_2}{r^4}\right),
\end{equation}
and
\begin{equation} \label{eq:3}
r^4_1=r^4_0\cosh^2\eta, \quad \textrm{and} \quad
r^4_2=r^4_0\sinh^2\eta .
\end{equation}

Our description of the baryon-like configurations will involve the
baryon vertex (the D5-brane), and various strings emanating from the
baryon vertex and ending on the asymptotic boundary or the black hole
horizon. The actions of the string worldsheets and the D5-brane can be
used to determine static solutions to the equations of motion. We will work below in the restricted context of the
AdS black hole described above, a more general formulation can be found
in \cite{A}.

In this section, we will illustrate the method by explicitly working out
the relevant physics of an $N_c=4$ baryon moving in the plasma. This
example is also our prime workhorse. The conflict between setting $N_c=4$
and the desire to have a large $N_c$ 
planar approximation where
finite string coupling effects are suppressed does not seem to be too much, because as we will see, the various results that were found in
\cite{A} can be reproduced in our case as well. In particular, in \cite{A}, large $N_c$ squashed baryons were considered, and the curves that we find later in this section are in excellent agreement with their work\footnote{The details of
the configurations we consider here are different from those considered
in \cite{A}, so this result also serves as a check of robustness for
the screening-length formula.}. So we believe that we are not throwing the baby out with the bath water by studying this simple system. Adding more quarks is in principle straightforward, except that the
computational effort is, of course, more.

The configuration we wish to study is shown in Figure 1. The
circular configuration of quarks lies in the $x_1$-$x_3$ plane and the
plasma wind is in the $x_3$ direction. In the figure we have suppressed
all coordinates except $r, x_1$ and $x_3$.
\begin{figure}[h]
\begin{center}
\includegraphics[width=0.8\textwidth
]{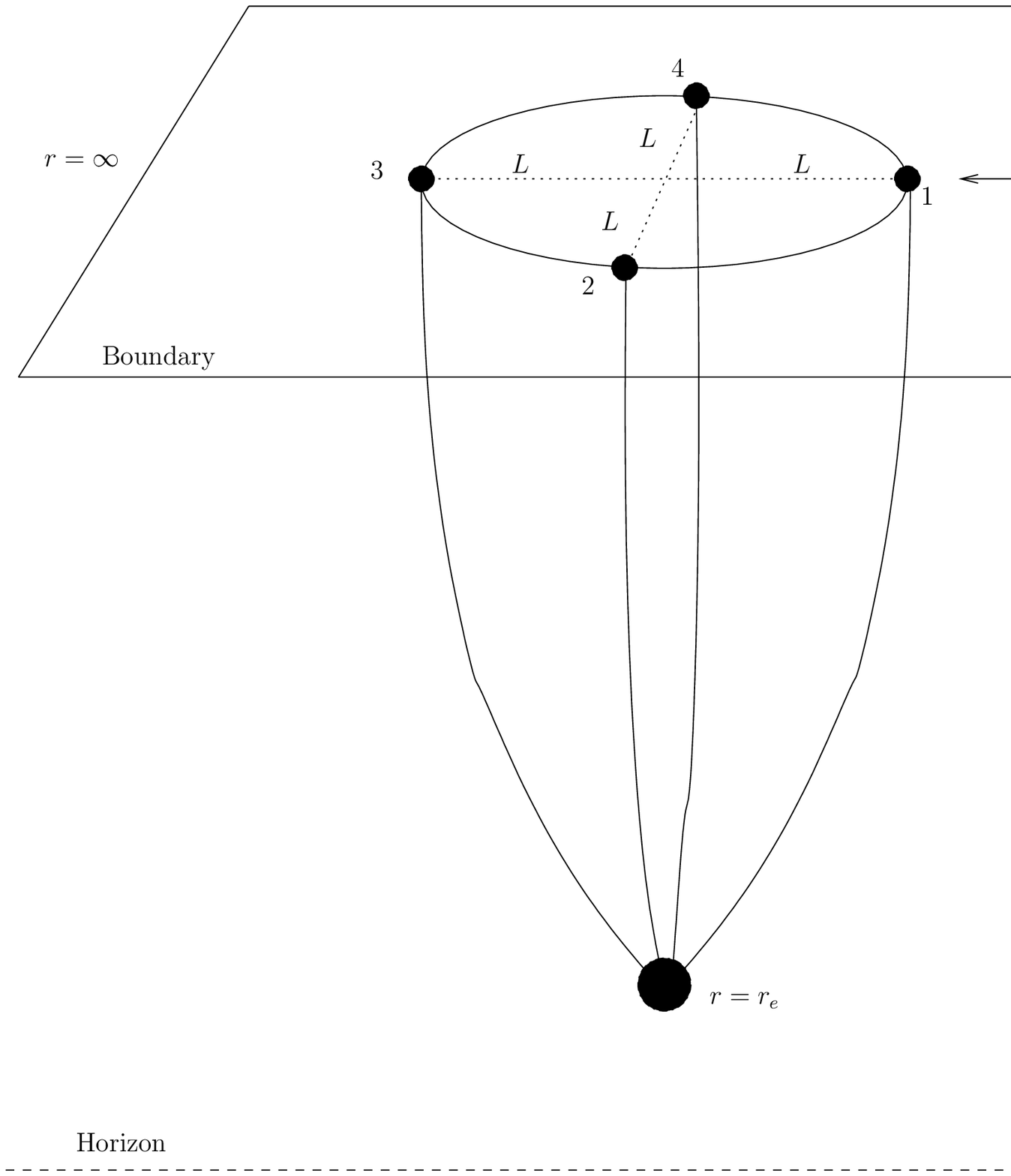}
\caption{
The $N_c=4$ baryon configuration in the AdS black hole background. The
baryon vertex is at $r=r_e$.
}
\label{Penrose}
\end{center}
\end{figure}
The action for
the system is given by
\beq
S=S_1+S_2+S_3+S_4+S_{D5},
\eeq
where subscripts denote the various strings and the D5-brane. The actions
for the various strings are computed \`a la Nambu-Goto in the
black hole background:
\beq
S_{NG}=\frac{1}{2\pi \alpha'}\int d \sigma d \tau \sqrt{-g_{\alpha\beta}},
\  {\rm with} \  g_{\alpha\beta}=G_{\mu\nu}\partial_\alpha
x^{\mu}\partial_\beta x^{\nu},
\eeq
where $G$ is the black hole metric and $g$ is the induced worldsheet
metric. If we assume that the configuration experiences no drag\footnote{
This assumption is supported by the computations involving meson
configurations \cite{Liu:2006nn, Liu:2006he, Peeters:2006iu,
chernicoff, Herzog, k-3} as
well as previously considered baryon configurations \cite{A, chernicoff2}. Our aim in this section is to set up the formalism and check whether we can connect with the results obtained in the literature, so we will not try to derive it {\em ab-initio}. We will assume no-drag for the baryon, and be content that the results work out precisely as expected. Of course, when we consider the dissociated configurations, we will not make the no-drag assumption.}, we can
look for axially symmetric circular configurations like the one shown in
Figure 1. For the two strings in the $x_1$ direction, then, we can take
the embedding to be
\beq
\tau=t, \ \ \ \sigma=r, \ \ \ x^{(a)}_1=x^{(a)}_1(\sigma)
\eeq
where $a$ is either  2 or 4 and denotes the appropriate string (see Figure
1). No $\tau$ dependence arises because we are interested in static
configurations. The action for the strings takes the form
\beq
S_a=\frac{{\cal T}}{2\pi \alpha'} \int_{r_e}^{\infty} dr
\sqrt{A\left(\frac{1}{f}+\frac{r^2}{R^2}\left({x^{(a)}_1}'\right)^2\right)},
\
\ \ a=2,4,
\eeq
where $\cal T$ can be thought of as the total time which gets divided out
in any relevant quantity. Primes denote derivatives with $r$. In the above
expression, $r_e$ is the position of
the baryon vertex and it can lie anywhere between $r=r_0$ and $r=\infty$,
i.e., between the boundary and the horizon. The boundary conditions for the
string coordinates are fixed by
the condition that the baryon vertex has coordinates $(r=r_e,x_1=0,
x_3=0)$. At the boundary, from a glance at the figure, we see that
$x^{(2)}_1(\infty)=-L$ and $x^{(4)}_1(\infty)=L$, where $L$ is the radius
of the
circle. For the strings in the $x_3$-directions, similarly, we get
\beq
S_b=\frac{{\cal T}}{2\pi \alpha'} \int_{r_e}^{\infty} dr
\sqrt{\left(\frac{A}{f}+\frac{r^2}{R^2}f\left({x^{(b)}_3}'\right)^2\right)},
\
\ \ b=1,3.
\eeq
The boundary conditions at infinity are given by
$x^{(1)}_3(\infty)=L,x^{(3)}_3(\infty)=-L$. The action for the D5-brane
vertex can
be taken as \cite{A}
\beq
S_{D5}=\frac{N_c R {\cal T}}{8\pi \alpha'}\sqrt{A(r_e)}.
\eeq

\subsection{Energetics and Screening Lengths}

The equations of motion for the configuration are obtained by varying
the total action with respect to the $x(r)$'s and also with respect to the
location of the
D5-brane. The details have been worked out in \cite{A} and the result
adapted to our case takes the following form. (We suppress the string
number superscripts $a(=2,4),b(=1,3)$ in some of the equations below.)\\
\underline{Strings 1, 3:}\\
\beq
\left(x_3'\right)^2=\frac{R^4}{f^2
r^2}\frac{K_3^2A}{\left(r^2f-R^2K_3^2\right)},
\ \ x_1' =0.
\eeq
The $K_3$ are integration constants. The $K_3$ have to satisfy the
condition,
\beq
K_3^{(1)}+K_3^{(3)}=0,
\eeq
as a force balance condition on the D5-brane. So we effectively need to
solve only for one of the strings.\\
\underline{Strings 2, 4:}\\
\beq
\left(x_1'\right)^2=\frac{R^4}{r^2}\frac{K_1^2}{f\left(r^2A-R^2K_1^2\right)},
\ \ x_3'=0.
\eeq
Again there is the relation
\beq
K_1^{(2)}+K_1^{(4)}=0
\eeq
that has to be satisfied.\\
\underline{D5-brane:}\\
\beq
\frac{2R\sqrt{A}}{\sqrt{f\left(R^2+fr^2(x_1')^2\right)}}\biggr|_{r=r_e}
+\frac{2RA}{\sqrt{f\left(AR^2+f^2r^2(x_3')^2\right)}}\biggr|_{r=r_e}=
\frac{r_e^4+r_1^4}{r_e^2\sqrt{r_e^4-r_1^4}}
\eeq
It should be noted that this equation is evaluated at $r=r_e$.
Since the string equations written down above depend only on the square of
the $K$'s, and these squares are identical for both strings in each pair,
we don't specify the superscript in the $x_1'$ and the $x_3'$ in the
D5-brane equation.

By introducing new variables, we can write these in the equivalent form
\beq
L=\frac{\rho \ \beta}{\pi T}\int_1^\infty dy
\frac{1}{(y^4-\rho^4)}\sqrt{\frac{y^4-\rho^4\cosh^2\eta}{y^4-\rho^4-\beta^2}},
\eeq
\beq
L=\frac{\rho \ \alpha}{\pi
T}\int_1^{\infty}\frac{dy}{\sqrt{(y^4-\rho^4)(y^4-\rho^4\cosh^2\eta-
\alpha^2)}},
\eeq
\beq
\frac{\sqrt{1-\rho^4\cosh^2 \eta}\sqrt{1-\rho^4- \beta^2}}{(1-\rho^4)}+
\frac{\sqrt{1-\alpha^2-\rho^4\cosh^2 \eta}}{\sqrt{1-\rho^4}}=\frac{1+\rho^4\cosh^2\eta}
{2\sqrt{1-\rho^4\cosh^2\eta}}.
\eeq
We have integrated the string equations of motion while imposing the
boundary condition that the radius of the circle is $L$ at the asymptotic
boundary. In the process we have also introduced the notation
\beq
\label{redef}
\alpha^2=\frac{K_1^2 R^4}{r_e^4}, \ \ \beta^2=\frac{K_3^2 R^4}{r_e^4}, \ \
y=\frac{r}{r_e}, \ \ \rho=\frac{r_0}{r_e}.
\eeq

To extract information about the baryon we need to solve these three equations
simultaneously. This can be done numerically.
\begin{itemize}
\item Pick a value for $\eta$ first.
\item Now, for each $\rho$ (it is easy to see that $\rho$ must lie
in the range $(0,1)$ from the geometry)
we can solve for $\beta$ in terms of $\alpha$ as we vary $\alpha$ (which again
has to be between $0$ and $1$.).
\item The correct value of $\alpha$ for each $\rho$, is the one where the
$\alpha$-integral matches the $\beta$-integral.
\item The value of the integral(s) at which this match happens is the
value of $L$.
\item Repeat the above procedure for another value of $\eta$.
\end{itemize}

Plot of $L$ as a function of $\rho$ are presented here,
for a few values of $\eta$. The plot for $L$ has a peak, and this is what one
identifies as the screening length, $L_s$. The results here match perfectly with those in \cite{A}.
\begin{figure}[h]
\begin{center}
\includegraphics[width=0.8\textwidth
]{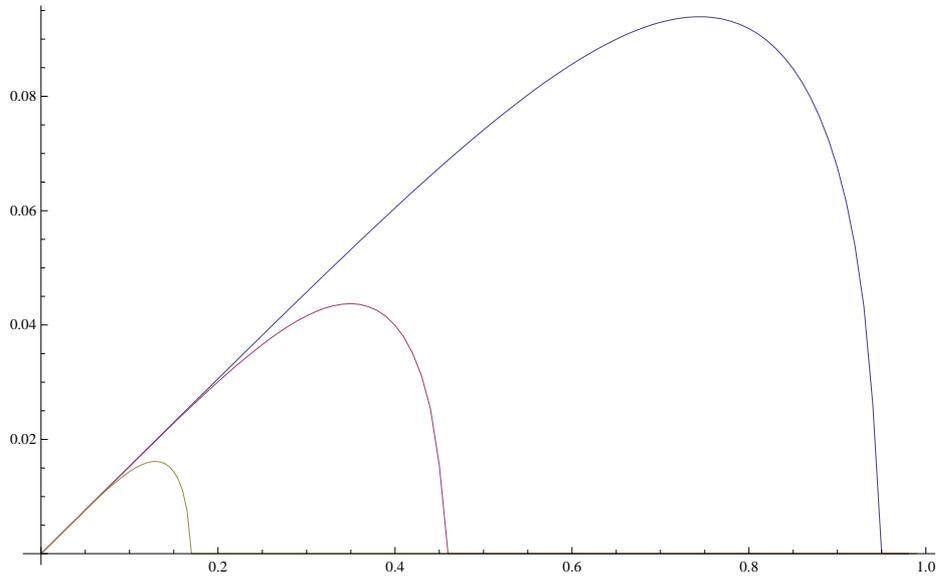}
\caption{$L T$ vs. $\rho$ for $\eta=0 \ ({\rm top \ curve}),2,4 \ ({\rm bottom \ curve})$. The maximum is associated to the screening length $L_s$ at the corresponding $\eta$.
}
\label{yyyy}
\end{center}
\end{figure}
\begin{figure}[h!]
\begin{center}
\includegraphics[width=0.8\textwidth
]{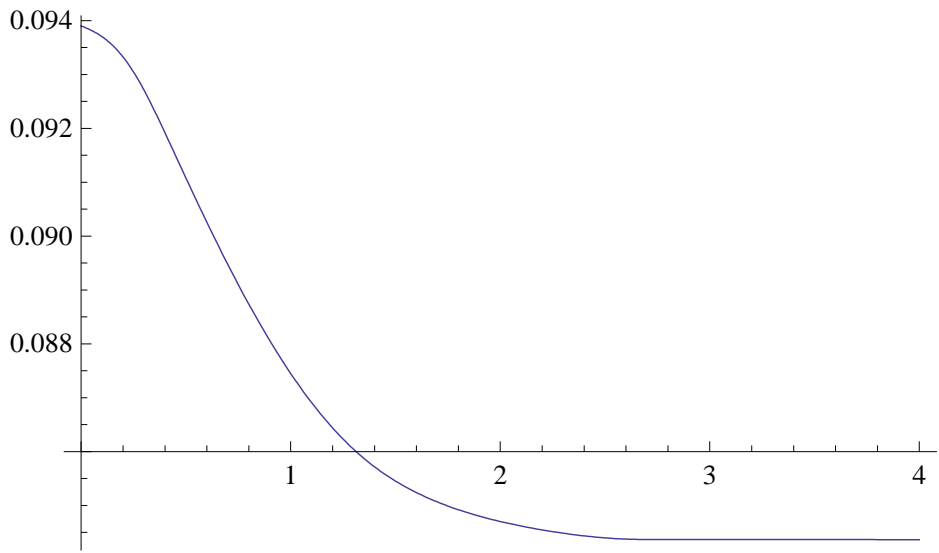}
\caption{$L_s T\sqrt{\cosh \eta}$ vs. $\eta$.
}
\label{Lscreen}
\end{center}
\end{figure}
The dependence of $L_s$ on the rapidity
$\eta$ can also been plotted, and the result again agrees with expectations
from other baryonic and mesonic configurations considered previously in the literature. In particular, it is evident from this plot that $L_s T \sim (1-v^2)^{1/4}$ for large boosts, and this is a check of the robustness of this result.

In the bulk of this paper, however, energetics are more interesting
to us than
screening lengths, so we turn to the computation of energy. The total
energy is the sum of the various pieces, with the caveat that the energy of
each string in the baryon
needs to be regulated by subtracting the energy of a free string
stretching all the way from the boundary to the horizon. Before the
regularization, the total energy is given by the (formal) expression
\begin{eqnarray}
E=\frac{1}{\pi \alpha'} \int_{r_e}^{\infty} dr
\sqrt{\frac{A}{f}+\frac{r^2f}{R^2}\left({x_3}'\right)^2}+
\frac{1}{\pi \alpha'} \int_{r_e}^{\infty} dr
\sqrt{\frac{A}{f}+\frac{r^2 A}{R^2}\left({x_1}'\right)^2}+\frac{R}
{2\pi \alpha'}\sqrt{A(r_e)}  \nonumber
\end{eqnarray}
Factors of two have been put in to take account of the two strings in each
pair. The last piece is the energy of the D5-brane.
The energy of the regulator  quark has been calculated in the literature,
we will use the expression (A.10) from \cite{Liu:2006he}. The idea is
to cut off the energy integral written above at $\Lambda$ instead
of $\infty$, subtract the quark energy integrated from the horizon to the
cutoff, and then take the limit $\Lambda$ to infinity {\em after} the
subtraction. The result is,
\begin{eqnarray}
E=T\sqrt{\lambda}\Biggl[\frac{1}{\rho}\int_1^{\infty}dy\left(\sqrt{
\frac{y^4-\rho^4\cosh^2\eta}{y^4-\rho^4-\beta^2}}-1\right)
+1-\frac{1}{\rho}+  \hspace{0.7in} \\
+\frac{1}{\rho}\int_1^{\infty}dy\left(
\frac{y^4-\rho^4\cosh^2\eta}{\sqrt{(y^4-\rho^4)
(y^4-\rho^4\cosh^2\eta-\alpha^2)}}-1\right)+1-\frac{1}{\rho}+
\frac{\sqrt{1-\rho^4\cosh^2\eta}}{2\rho}\Biggr].\nonumber
\end{eqnarray}
where we have used the variables \ref{redef} as well as
Maldacena's relations and the geometrical definition of the Hawking temperature.

\begin{figure}[h!]
\begin{center}
\includegraphics[width=0.8\textwidth
]{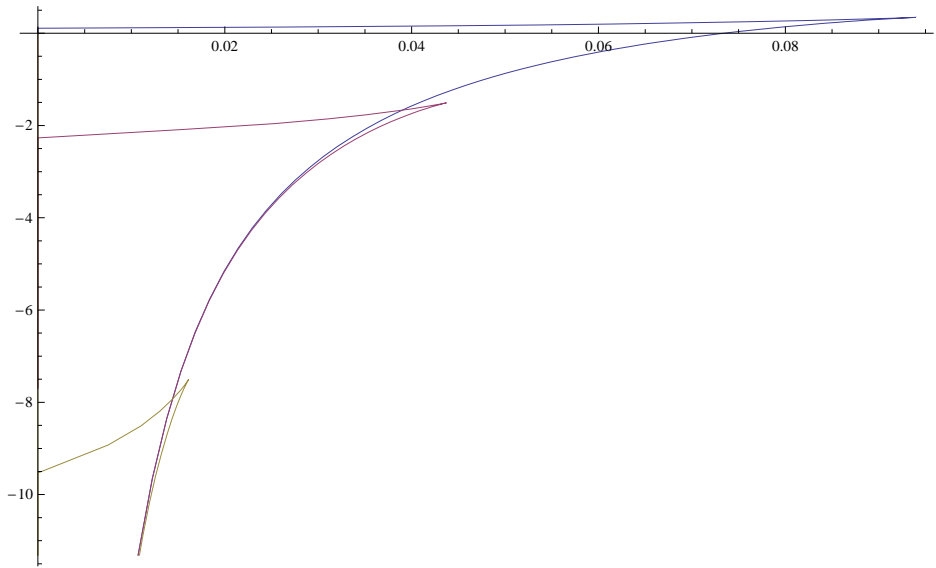}
\caption{
$\frac{E}{T\sqrt{\lambda}}$ vs. $L$, for $\eta=0 \ ({\rm top \ curve}),2,4 \ ({\rm bottom \ curve})$.
}
\label{xxxx}
\end{center}
\end{figure}

With the values of $L$, $\alpha$, $\beta$ that we computed earlier, it is
possible to make a numerical plot between $L$ and $E$ for various values
of $\eta$ and we present the results in the figure. This ties in as it should, with the similar plot in \cite{A}.

The computations of this section serve two purposes. Firstly and primarily for us, they give a context for the rest of the work on this paper. But in the process, they also provide a confirmation of the robustness of previous results on baryons. In \cite{A}, qualitatively identical results were observed, where instead of keeping the baryons on a circle, the angle at which the strings hit the D5-brane was fixed. This amounts to considering quark configurations that are squashed in the plasma wind. This makes the computations technically different, but still we see from a glance at the plots that the essential features are identical and therefore that the results are indeed very robust. In particular, the screening length formula $L_s T \sqrt{\cosh \eta} \sim {\rm const.}$ holds in our case as well for large $\eta$, even though the value of the constant seems slightly different\footnote{The slight variation in the constant is not bad - its value is known to depend on the details of the configuration, see e.g.  figure 7, in \cite{A}.}.

\section{Dissociated Baryons}

Clearly there are many possible ways in which the baryon presented in the
previous section can dissociate. One possibility is to look at configurations where one quark has dissociated, while keeping as many of the remaining quarks as possible still on the circle. After the force balance conditions are imposed on the D5-brane, there are only a few interesting configurations which one can consider in this manner (we will look at them in Appendix C). But since the process of ripping a quark is fundamentally dynamical, it is not clear to us that such configurations where the quarks are forced to be in a circle are the only ones preferred. Since it is difficult to come up with an unambiguous definition of what is ``closest" and what is ``farthest" to the original baryonic configuration\footnote{Also, it is not clear how general our understanding would be even if we were to come up with such a notion for the four-quark baryon. }, we will attempt something more modest here. We will instead consider two baryonic configurations with identical linear dimensions but different orientations. We hope to make some comments about their relative energetics from this, but notice that we are steering clear of the question of what are the preferred dissociation products.

The two configurations we consider are as follows. One is when the undissociated quarks are in a rigid identically spaced line parallel to the wind (i.e., along $x_3$)
(case I), and the other is when they are in the direction perpendicular
to the wind (i,e., along $x_1$) (case II). We will also have a dissociated string in each case, which we will
take to be well-separated from the rest of the baryon. Our aim is to look at two  configurations that are identical, except for their orientation in the wind. Even though we will need to work much harder to get a fuller understanding of phenomenology, to get a hint about the orientation-dependence of the energy, this computation should be enough.

We address both the cases separately. But before doing so, we notice a useful fact: the dissociated string that trails all the way to the horizon from the D5-brane is forced to lie
along $x_3$. This is demonstrated in Appendix B.

\subsection{Case I: Longitudinal Quarks}

\begin{figure}[h!]
\begin{center}
\includegraphics[width=0.5\textwidth
]{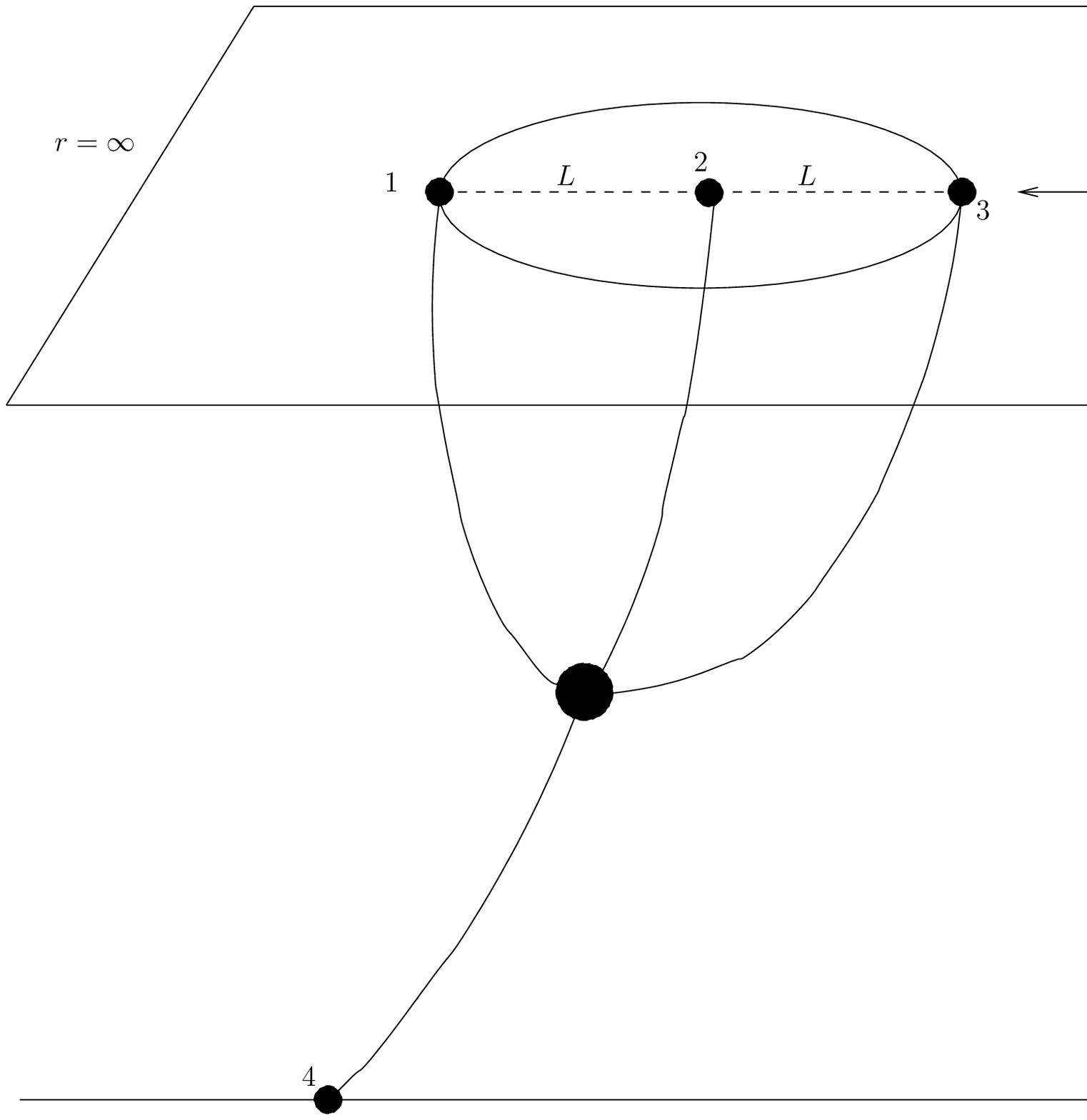}
\caption{Case I: Remnant with quarks along the plasma wind.
}
\label{longi}
\end{center}
\end{figure}

We first consider the case where the undissociated quarks
are along the direction of the wind. After
the dissociation, the string corresponding to the dissociated
quark is trailing, with one of its endpoints at the D5-brane vertex and
the other at the horizon. The configuration we are considering is demonstrated in figure \ref{longi},
and consists of three quarks in a straight-line. In case II, we will consider
three quarks, again in a straight line, but perpendicular to the wind. The hope is that
these two complimentary configurations will give us some idea about the energetics. Notice that because of the constraints arising from force balance at the D5-brane and because the trailing string is forced to lie along $x_3$, many of the configurations are ruled out. An advantage of comparing the two cases we consider here is that they give a natural way to contrast {\em between} the cases: we can compare quark configurations of equal linear dimensions on either side. This is useful in getting some insight about how the energy of a given configuration (with fixed linear dimensions) changes with the orientation.

\noindent

We can compute the energy of this configuration using methods
similar to those of last section. In what follows, $\beta_i, \alpha_i$ are defined as in the last section (the indices in the present case denote the respective quark). In case I, the $\alpha_i$ are identically zero. The origin of the coordinate system is taken to be at the D5 brane ($x_1(D5)=x_3(D5)=0$).
The numbering of the quarks are indicated in the figure. The trailing string solution forces the relation
\bea
\beta_4=\rho^2 \sinh \eta.
\eea
Using the results of Appendix A, we can write down the vertical force balance condition at the D5-brane as
\bea
\frac{\sqrt{1-\rho^4 \cosh^2 \eta }}{1-\rho^4}\left(\sum_{i=1}^{3}\sqrt{1-\rho^4-\beta_i^2}-\sqrt{1-\rho^4-\beta_4^2}\right)=
\frac{1+\rho^4 \cosh^2 \eta}{\sqrt{1-\rho^4 \cosh^2 \eta}}, \label{d1v}
\eea
and the horizontal balance as,
\bea
\beta_1+\beta_2+\beta_3=\beta_4 (=\rho^2 \sinh \eta). \label{d1h}
\eea
In case I, we will look at configurations with the three undissociated quarks placed equidistantly on a line, along the wind. 
 Thus we have the two equations below for the distance between quarks.
\bea
L_{12}\equiv L=\frac{\rho}{\pi T}\int_1^\infty dy
\frac{\sqrt{y^4-\rho^4\cosh^2\eta}}{(y^4-\rho^4)}\left(\frac{\beta_2}{\sqrt{y^4-\rho^4-\beta_2^2}}-\frac{\beta_1}{\sqrt{y^4-\rho^4-\beta_1^2}}\right),
\nonumber \\
L_{23}\equiv L=\frac{\rho}{\pi T}\int_1^\infty dy
\frac{\sqrt{y^4-\rho^4\cosh^2\eta}}{(y^4-\rho^4)}\left(\frac{\beta_3}{\sqrt{y^4-\rho^4-\beta_3^2}}-\frac{\beta_2}{\sqrt{y^4-\rho^4-\beta_2^2}}\right).
\nonumber
\eea

\noindent
These equations follow directly when integrating the equations of motion presented in Appendix B, for the specific case under consideration here.

An algorithm for solving the system is as follows. For each value of $\eta$ we do the following:
\begin{itemize}
\item Pick a value of $\rho$.
\item For each value of $\rho$, scan $\beta_1$ between -1 and 1.
\item For each value of $\beta_1$, solve (\ref{d1v}) and (\ref{d1h}) simultaneously to obtain $\beta_2, \beta_3$, paying attention to the signs using the geometry of the configuration.
\item Now we can compute $L_{12}$ and $L_{23}$ for each value of $\beta_1$. The value of $\beta_1$ at which the two coincide is the correct value of $\beta_1$, and the corresponding $L$ can be plotted as a function of $\rho$ (for each given $\eta$). Once we have fixed $\beta_1$, it is straightforward to fix $\beta_2, \beta_3$ using (\ref{d1v}) and (\ref{d1h}).
\end{itemize}
This can be repeated for each value of $\eta$. Once we have $L$ and $\beta_i$ as functions of $\rho$, we can compute the energy of the configuration, which is what we are really after. To compare the energies of cases I and II, we will not need to worry about the far-separated quark, so the regulated energy of the configuration shown in the figure takes the form:
\bea
E=T\sqrt{\lambda}\Biggl[\frac{1}{2\rho}\int_{1}^{\infty}dy \sum_{i=1}^{3}\bigg(\sqrt{
\frac{y^4-\rho^4\cosh^2\eta}{y^4-\rho^4-\beta_i^2}}-1 \bigg) +1-\frac{1}{\rho}+\frac{\sqrt{1-\rho^4\cosh^2\eta}}{2\rho}\Biggr].
\eea
Here and in the next subsection, we have subtracted the energy of three regulator strings as opposed to four in the case of the undissociated baryon.
We will present plot of $L$ vs. $\rho$ and $E$ vs. $L$ for both cases I and II together at the end of the next subsection.

\subsection{Case II: Transverse Quarks}

\begin{figure}[h]
\begin{center}
\includegraphics[width=0.5\textwidth
]{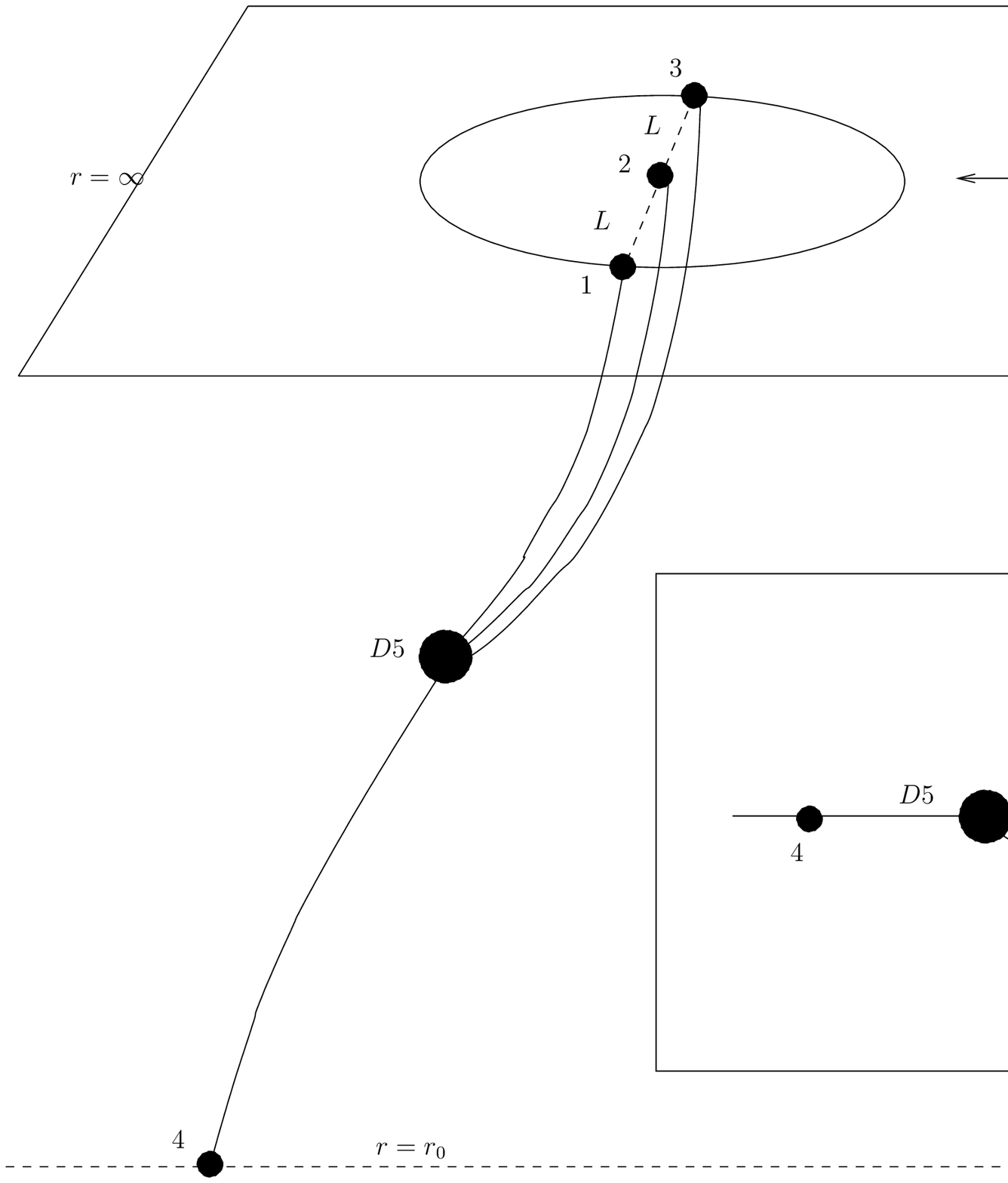}
\caption{Case II: Remnant with transverse quarks (Box shows top view).
}
\label{transverse}
\end{center}
\end{figure}

Case II corresponds to (undissociated) quarks aligned perpendicular to the wind as in figure \ref{transverse}. The Nambu-Goto Lagrangian for the string $i$ that is stretched both in the $x_3$ and $x_1$ directions is given by
\bea
S_i=\frac{{\cal T}}{2 \pi \alpha'}\int_{r_e}^{\infty}dr \sqrt{A\Big(\frac{1}{f}+\frac{r^2\big(x_1^{(i)}\big)'^2}{R^2}\Big)+\frac{r^2f(r)\big(x_3^{(i)}\big)'^2}{R^2}}.
\eea
The energies, equations of motions etc. for quarks 1 and 3 in figure \ref{transverse} are computed with this expression.

Again, from the generic equations of motion written down in Appendix A, we find that the horizontal force balance conditions at the D5-brane enforces
\bea
|\alpha_3|=|\alpha_1|\equiv \alpha, \ \ \beta_3=\beta_1\equiv \beta, \\
\beta_2+2\beta=\beta_4 (=\rho^2 \sinh \eta ). \label{d2h}
\eea
The vertical balance condition can be written as
\bea
\frac{(1-\rho^4 \cosh^2 \eta)}{(1-\rho^4)}+\frac{1+\rho^4 \cosh^2 \eta}{\sqrt{1-\rho^4 \cosh^2 \eta}} = \hspace{1.5in}\nonumber \\
\frac{2\sqrt{(1-\rho^4 \cosh^2 \eta)(1-\rho^4-\beta^2)-(1-\rho^4)\alpha^2}}{(1-\rho^4)}+
\frac{\sqrt{(1-\rho^4 \cosh^2 \eta)(1-\rho^4-\beta_2^2)}}{(1-\rho^4)} \label{d2v}
\eea
The quarks 1, 2 and 3 all have the same $x_3$ coordinate because of the configuration we have chosen. This means that
\bea
\frac{\rho \ \beta_2}{\pi T}\int_1^\infty dy
\frac{1}{(y^4-\rho^4)}\sqrt{\frac{y^4-\rho^4\cosh^2\eta}{y^4-\rho^4-\beta_2^2}} = \hspace{1.3in} \label{x3}\\
=\frac{\rho \beta}{\pi T}\int_1^\infty dy\frac{(y^4-\rho^4\cosh^2\eta)}{(y^4-\rho^4)\sqrt{(y^4-\rho^4)(y^4-\rho^4\cosh^2\eta)-\beta^2(y^4-\rho^4\cosh^2\eta)-\alpha^2(y^4-\rho^4)}}. \nonumber
\eea

\noindent
The spacing between the undissociated quarks (fixed to be $L$) is given by their $x_1$ coordinate, which can be obtained by integrating the equation of motion:
\bea
L=\frac{\rho \alpha}{\pi T}\int_1^\infty dy\frac{1}{\sqrt{(y^4-\rho^4)(y^4-\rho^4\cosh^2\eta)-\beta^2(y^4-\rho^4\cosh^2\eta)-\alpha^2(y^4-\rho^4)}}.
\label{d2L}
\eea
These equations can again be solved numerically for any value of $\eta$:
\begin{itemize}
\item Pick a value of $\rho$.
\item For each value of $\rho$, scan $\beta_2$ between 0 and 1.
\item For each value of $\beta_2$, solve (\ref{d2h}) and (\ref{d2v}) simultaneously to obtain $\alpha, \beta$.
\item Now we can evaluate the RHS and LHS of (\ref{x3}) for each value of $\beta_2$. The value of $\beta_2$ at which the two coincide is the correct value of $\beta_2$. Once we have fixed $\beta_2$, it is straightforward to fix $\alpha, \beta$ using (\ref{d2h}) and (\ref{d2v}). The corresponding $L$ can be plotted as a function of $\rho$ (for each given $\eta$) using (\ref{d2L}).
\end{itemize}

The regulated energy of the dissociated case II configuration can be calculated similar to the previous cases (here again, we subtract three quarks as in case I) and the result is
\bea
E=T\sqrt{\lambda}\Biggl[ \frac{1}{\rho}\int_{1}^{\infty}dy \bigg(\frac{(y^4-\rho^4\cosh^2\eta)}{\sqrt{(y^4-\rho^4\cosh^2\eta)(y^4-\rho^4-\beta^2)-\alpha^2(y^4-\rho^4)}}-1\bigg)+\nonumber \\
+\frac{1}{2\rho}\int_{1}^{\infty}dy \bigg(\sqrt{
\frac{y^4-\rho^4\cosh^2\eta}{y^4-\rho^4-\beta_2^2}}-1 \bigg)+1-\frac{1}{\rho}+\frac{\sqrt{1-\rho^4\cosh^2\eta}}{2\rho}\Biggr].
\eea

\subsection{Plots}

\begin{figure}[h!]
\begin{center}
\includegraphics[width=0.8\textwidth
]{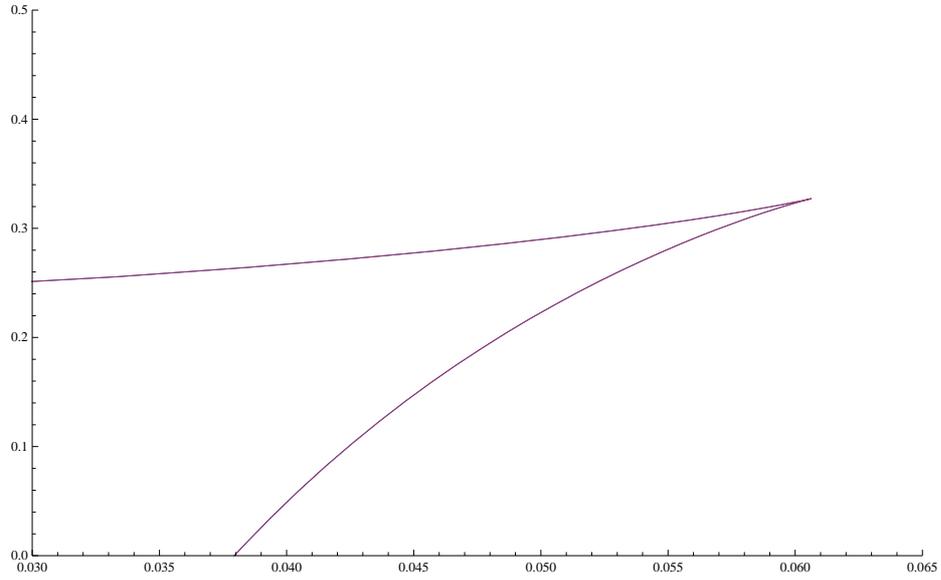}
\caption{The $\frac{E}{T\sqrt{\lambda}}$ vs. $L$ plots of the transverse and longitudinal configurations at $\eta=0$. Even though the numerics that gives rise to the curves is different, the two are exactly on top of each other.
}
\label{en0}
\end{center}
\end{figure}
\begin{figure}[h!]
\begin{center}
\includegraphics[width=0.8\textwidth
]{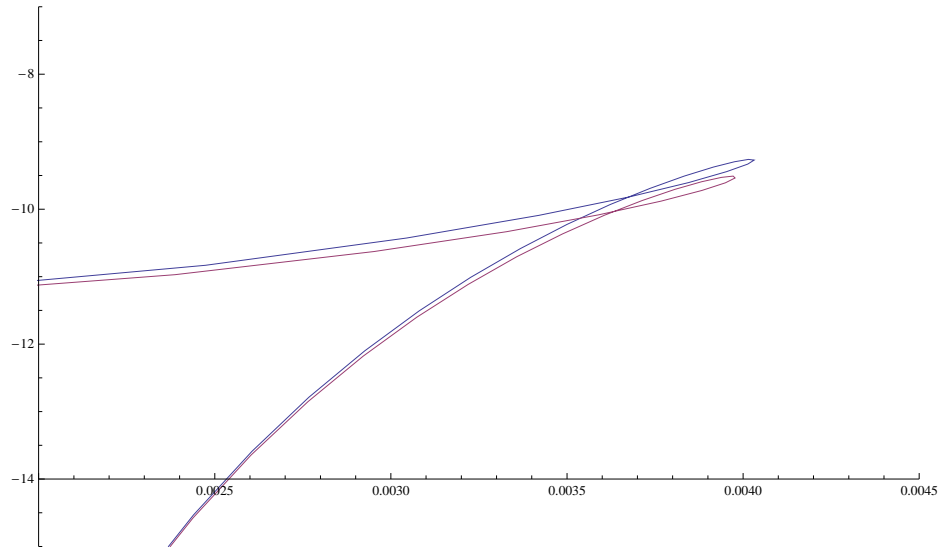}
\caption{The $\frac{E}{T\sqrt{\lambda}}$ vs. $L$ plots for the two cases at $\eta=6$. The longitudinal curve is the one at higher energy. The qualitative features are identical at other values of $\eta \neq 0$.
}
\label{en6}
\end{center}
\end{figure}

\begin{figure}[h!]
\begin{center}
\includegraphics[width=0.8\textwidth
]{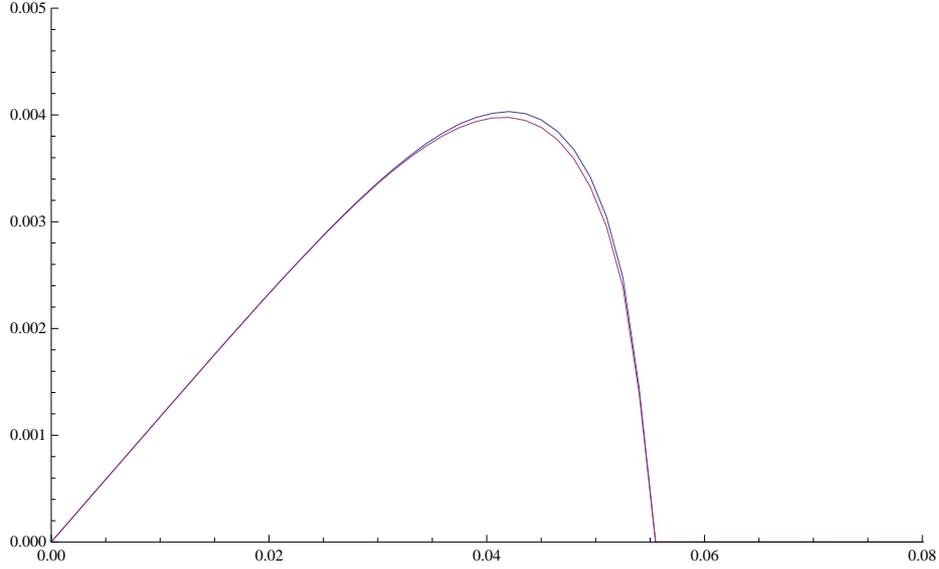}
\caption{$LT$ vs. $\rho$ for $\eta=6$ for the two cases. The transverse case (II) is marginally lower, as expected from the intuition of figure 4 in \cite{A}.
}
\label{screen-diss-6}
\end{center}
\end{figure}

\begin{figure}[h!]
  \begin{center}
    \subfloat[$\eta$=0.]{\label{zerobar}\includegraphics[scale=0.55]{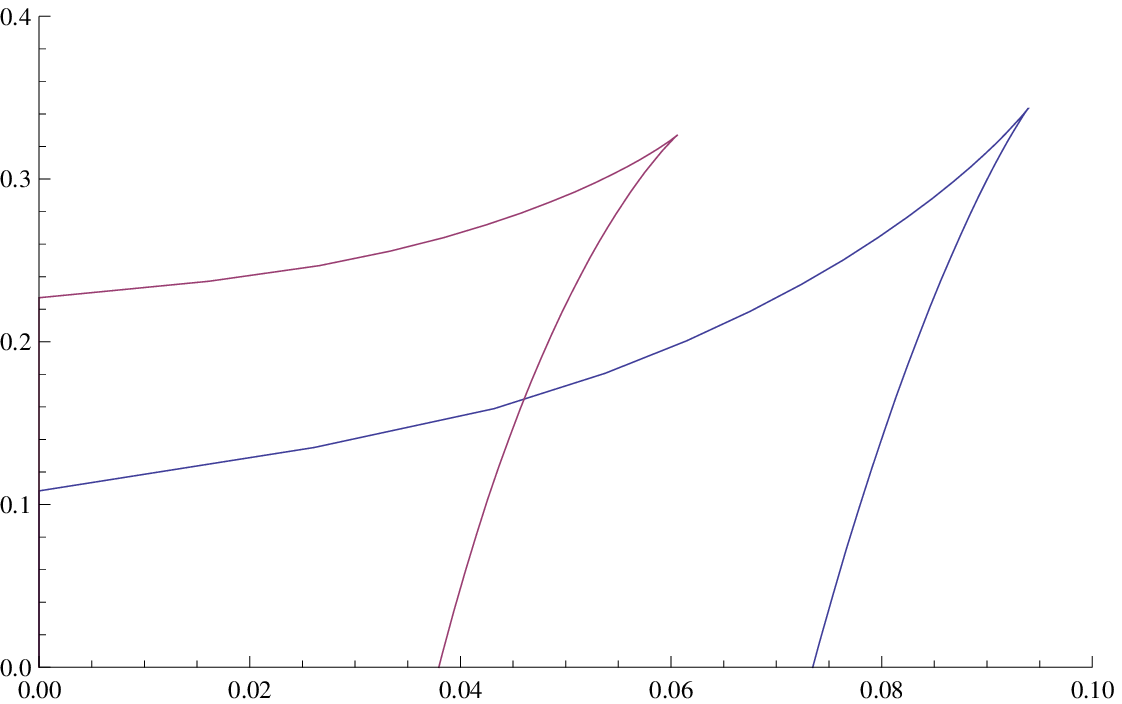}}
    \subfloat[$\eta=6$]{\label{barvsdiss}\includegraphics[scale=0.45]{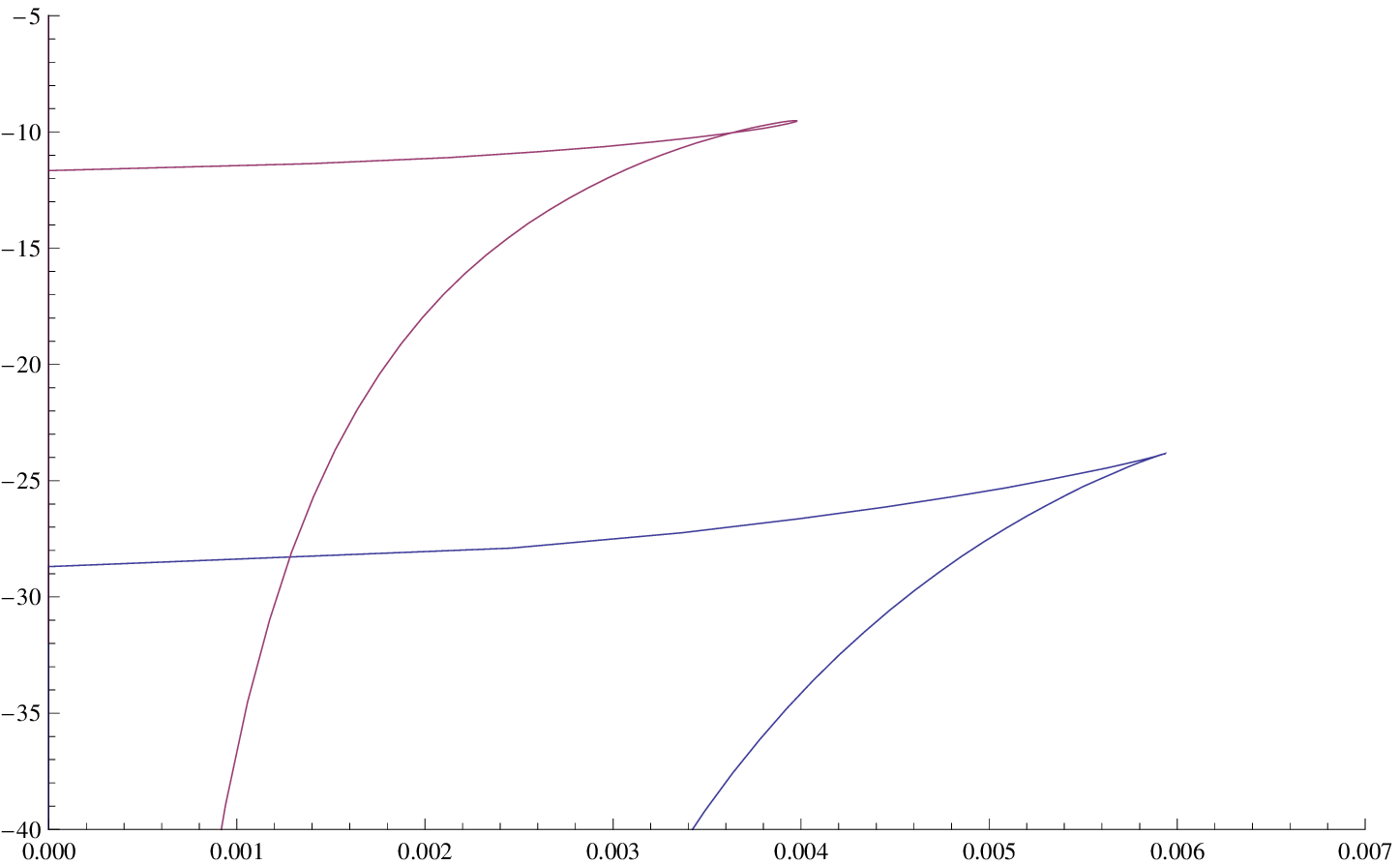}} \\
  \end{center}
  \caption{Plots of $\frac{E}{T\sqrt{\lambda}}$ vs. $LT$ for baryons (lower curves) and dissociated configurations (upper curves). For less clutter we only show the transverse case.}
  \label{panels}
\end{figure}
Using the expressions from the previous subsections and the numerical simulations resulting from them, we can
make comparisons between the energies of the two dissociated configurations.
The curves are similar in both cases, except that the longitudinal quarks are at a higher energy when the windspeed is non-zero. The energy plots should be identical when the windspeed goes to zero since case I and II are identical in this case, so this can be used as a consistency check of our numerics. This is indeed what we find as clear from the $\eta=0$ plot. We show the plots of $E$ vs. $L$  for two representative cases ($\eta=0$ and $\eta=6$), in their interesting regimes of parameters. The fact that longitudinal configurations are at a higher energy than transverse ones has previously been observed in the case of mesonic configurations in \cite{Liu:2006he} (see right panel of figure 6 in \cite{Liu:2006he}).

Another observation one can make from the plots is that the energies of the dissociated configurations are always above those of the corresponding undissociated configurations for sufficiently large values of $\eta$, and this is shown in Figure \ref{barvsdiss}. This result is reasonable. The zero of the energy at each $\eta$ is set by the (colored) trailing string at {\em that} $\eta$. We also know from a previous section (see also \cite{A}) that baryons have negative energy with respect to this datum, for sufficiently large windspeeds. So the configurations we are investigating here, which are morally between baryons and free strings, should naturally have intermediate energies. For small enough windspeed, there is some changes in these comparative plots, and we present them in Figure \ref{zerobar}. The cusp region is above zero energy even for the undissociated baryon in this case, as expected (see also \cite{A}).

The structure of the energy curves for the dissociated cases is roughly 
\begin{wrapfigure}{r}{60mm}
  \begin{center}
    \includegraphics[height=0.3\textheight]{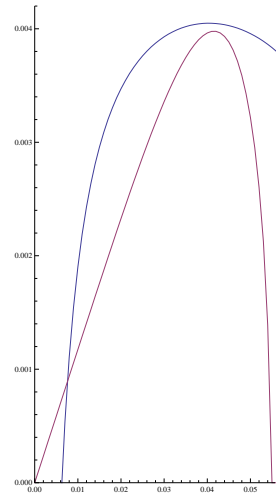}
  \end{center}
  \caption{Typical shapes of $E$ and $L$ vs. $\rho$ for any $\eta$ and for any configuration. The axes have been re-scaled to fit the curves on the same plot, and the specific values are not to be paid attention to.}
  \vspace{-0.1in}
\end{wrapfigure}
similar to that of the undissociated case, but in the plots here, we have zoomed in on the fine structure. The one significant difference with the baryonic case is that instead of a cusp, for non-zero boost, here we  see a loop. We have done these numerical simulations for various configurations and various resolutions (see also Appendix C), so it seems very unlikely that this loop is a result of some systematic numerical error.
The loops arise as a robust feature of all dragging configurations. In particular, they vanish for color non-singlet configurations only when the windspeed is zero and there is no drag (see Figure \ref{en0}). The technical reason why the loops arise in E vs. L plots is not hard to see. The schematic plots of $E$ vs. $\rho$ and $L$ vs. $\rho$ for generic configurations and generic $\eta$ are given in the figure to the right. They both have one maximum, and for the case of non-drag configurations, the peaks in both $E$ and $L$ happen for the same value of $\rho$. But for dragging configurations, the peaks in $E$ are displaced to the left with respect to the peak in $L$, and therein lie the origins of the loop. It would be nice to understand the physical origins of this shift, better.

We suspect that these loops can be used as a measure of the drag of the configuration, because their areas seem to vary depending on the configurations\footnote{For the cases considered in this section, this effect is not pronounced.}. The fact that the loops have non-zero area for dragging configurations, and the fact that the area of the loop in the $\frac{E}{T\sqrt{\lambda}}$ vs. $LT$ plot has dimensions of energy in string units, suggests that this could perhaps be used as a measure of the energy loss. It would certainly be very interesting to study this further, and we plan to come back to this in the future.

\section{Discussions and Loose Ends}

We have already reported the main conclusions in the introduction, so here we will merely make some comments.

There are some natural extensions to the work done here. We have considered the special case of $N_c=4$ quarks, and found that we can reproduce and extend the results in the literature that deal with color singlet configurations. Considering a uniform distribution of quarks along the circle where the large $N_c$ limit is more systematic would be a natural next step. This will be more complicated because the strings are now not along the coordinate axes. More generic gauge theories, configurations other than circular
baryons, more generic wind directions etc. are all possible lines of investigation, especially in understanding the possibility of extracting universal predictions valid across conformal/confining gauge
theories.

In this paper, we have compared dissociated configurations which capture some aspects of the energetics of baryon dissociation. But it should be emphasized that the results of this paper are tentative. It would be nice to investigate other configurations, especially those with more quarks. Another interesting possibility is to do a scan of the various dissociation configurations at various angles and lengths. In particular, investigating other configurations which are identical except for their orientations would be useful\footnote{Not all such configurations are kinematically allowed.}.

Our original configuration of four quarks was specifically chosen to simplify the calculations. This is a perfectly good starting point, but it also suffers from the drawback that we cannot investigate many interesting configurations. A good example is the transverse and longitudinal configurations (analogous to the two configurations we investigated in section 3) after two quarks have dissociated. Both these configurations are not allowed because the force balance conditions on the D5-brane are too restrictive.

To get a full and unambiguous understanding of the possible dissociation patterns, we need a more exhaustive study of the various hints we have found in this paper. We have tried to stay close to the configurations which are most easily tractable. Our aim here has been to set up some of the framework and leave the more thorough work to a more elaborate future project, perhaps with more man-power. Some of these questions are currently under investigation.



\section{Acknowledgments}

It is a pleasure to thank Krishna Rajagopal for encouragement, clarifying correspondence and helpful comments on a previous version of the manuscript. I also  thank Christina Athanasiou, Francesco Bigazzi, Stanislav Kuperstein and Carlo Maccaferri for discussions. This work is supported in part by IISN - Belgium (convention 4.4505.86), by the Belgian National Lottery, by the
European Commission FP6 RTN programme MRTN-CT-2004-005104 in which the
author is associated with V. U. Brussel, and by the Belgian Federal
Science Policy Office through the Interuniversity Attraction Pole P5/27.

\section*{\bf Appendix}
\addcontentsline{toc}{section}{Appendix}


\subsection*{{\bf A.} \ Equations of Motion}
\addcontentsline{toc}{subsection}{{\bf A} \ Equations of Motion}
\renewcommand{\theequation}{A.\arabic{equation}}

The equations governing the various static baryon-like
configurations is obtained by varying the total action with respect to
the string coordinates and with respect to the
position of the D5-brane. Since the boundaries of the strings (at the
D5-brane) are also supposed to be varying, one ends up getting a bit more
than the usual Euler-Lagrange
equations. The basic ideas are presented in \cite{A}, but we have chosen
to redo it here for the case when there are
dragging strings to clarify the origin of some negative signs which turn
out to be crucial in this work.

The action for the system takes the general form
\beq
S=\sum_{m\in \{{\rm up}\}}\int_{r_e}^{\infty}{\cal
L}_m(x_{i,m}',r)dr+\sum_{n\in \{{\rm down}\}}\int_{r_0}^{r_e}{\cal
L}_n(x_{i,n}',r)dr+S_{D5}(r_e),
\eeq
where we have denoted all the functional dependencies on the relevant
variables. $x_{i,m}$ stands for the the $i$-th coordinate of string $m$.
Primes are, again, with respect to $r$. The summation over the
strings will be written explicitly in what follows, but the summation
over $i$, should be understood from the context.

First we vary with respect to the $x$'s and get the individual equations
of motion for the various strings. But since the boundaries are also
allowed to vary, we also get boundary equations of motion which give
further constraints on the configurations. Setting $\delta S=0$ (for
variations of $x_i$), we end up with
\begin{eqnarray}
0=\sum_{m\in \{{\rm up}\}}\int_{r_e}^{\infty}dr
\frac{\partial{\cal L}_m}{\partial x_{i,m}'}\delta x_{i,m}'
+\sum_{n\in \{{\rm down}\}}\int_{r_0}^{r_e}dr
\frac{\partial{\cal L}_n}{\partial x_{i,n}'}\delta
x_{i,n}' \hspace{0.9in}\\
= \sum_{m\in \{{\rm up}\}}\int_{r_e}^{\infty}dr
\biggl[\frac{d}{dr}\left(\frac{\partial{\cal
L}_m}{\partial x_{i,m}'}\delta
x_{i,m}\right)-\frac{d}{dr}\left(\frac{\partial{\cal L}_m}{\partial
x_{i,m}'}\right)\delta x_{i,m} \biggr]+\sum_{n\in \{{\rm
down}\}}\int_{r_0}^{r_e}dr\biggr[m \rightarrow n\biggl]. \nonumber
\end{eqnarray}
We have written $\delta x_{i,m}'$ as $d(\delta x_{i,m})/dr$ and done an
integration by parts as usual. Since the variations $\delta x_i$ are
arbitrary in the bulk (of the string), the second
term in each piece has to be zero and gives rise to the standard
Euler-Lagrange equations. In the present case they take the form:
\begin{eqnarray}
\frac{\partial{\cal L}_m}{\partial
x_{i,m}'}={\rm const.}\equiv K_{i,m}, \ \ \frac{\partial{\cal
L}_n}{\partial
x_{i,n}'}={\rm const.}\equiv K_{i,n}.
\end{eqnarray}
What remains in the variation is
\begin{eqnarray}
0=\sum_{m\in \{{\rm up}\}}\int_{r_e}^{\infty}dr
\frac{d}{dr}\left(\frac{\partial{\cal
L}_m}{\partial x_{i,m}'}\delta x_{i,m}\right)+\sum_{n\in \{{\rm
down}\}}\int_{r_0}^{r_e}dr \frac{d}{dr}\left(\frac{\partial{\cal
L}_n}{\partial x_{i,n}'}\delta x_{i,n}\right) \nonumber \\
= -\sum_{m\in \{{\rm up}\}}\frac{\partial{\cal
L}_m}{\partial x_{i,m}'}\delta x_{i,m}\bigg|_{r=r_e} + \sum_{n\in \{{\rm
down}\}} \frac{\partial{\cal
L}_n}{\partial x_{i,n}'}\delta x_{i,n}\bigg|_{r=r_e} \label{boundy}
\hspace{+0.5in}
\end{eqnarray}
The strings are fixed at $r=r_0$ and $r=\infty$, so the terms from
those ends of the integral don't
contribute. Also, at $r=r_e$, the strings are allowed to move, but only
under the constraint that they are all still attached at the baryon
vertex. This means that $\delta x_{i,m}|_{r_e}=\delta
x_{i,n}|_{r_e}\equiv\delta x_{i}|_{r_e}$. Another input comes from
the Euler-Lagrange equations above. They say, in particular, that
$\frac{\partial{\cal
L}}{\partial x_{i}'}\big|_{r_e}=K_i$.
Putting this all together,
we end up with
\beq
\sum_{m\in \{{\rm up}\}}K_{i,m}-\sum_{n\in \{{\rm down}\}}K_{i,n}=0,
\eeq
which gives rise to three equations, one for each $i$.

\begin{figure}[h]
\begin{center}
\includegraphics[width=0.8\textwidth
]{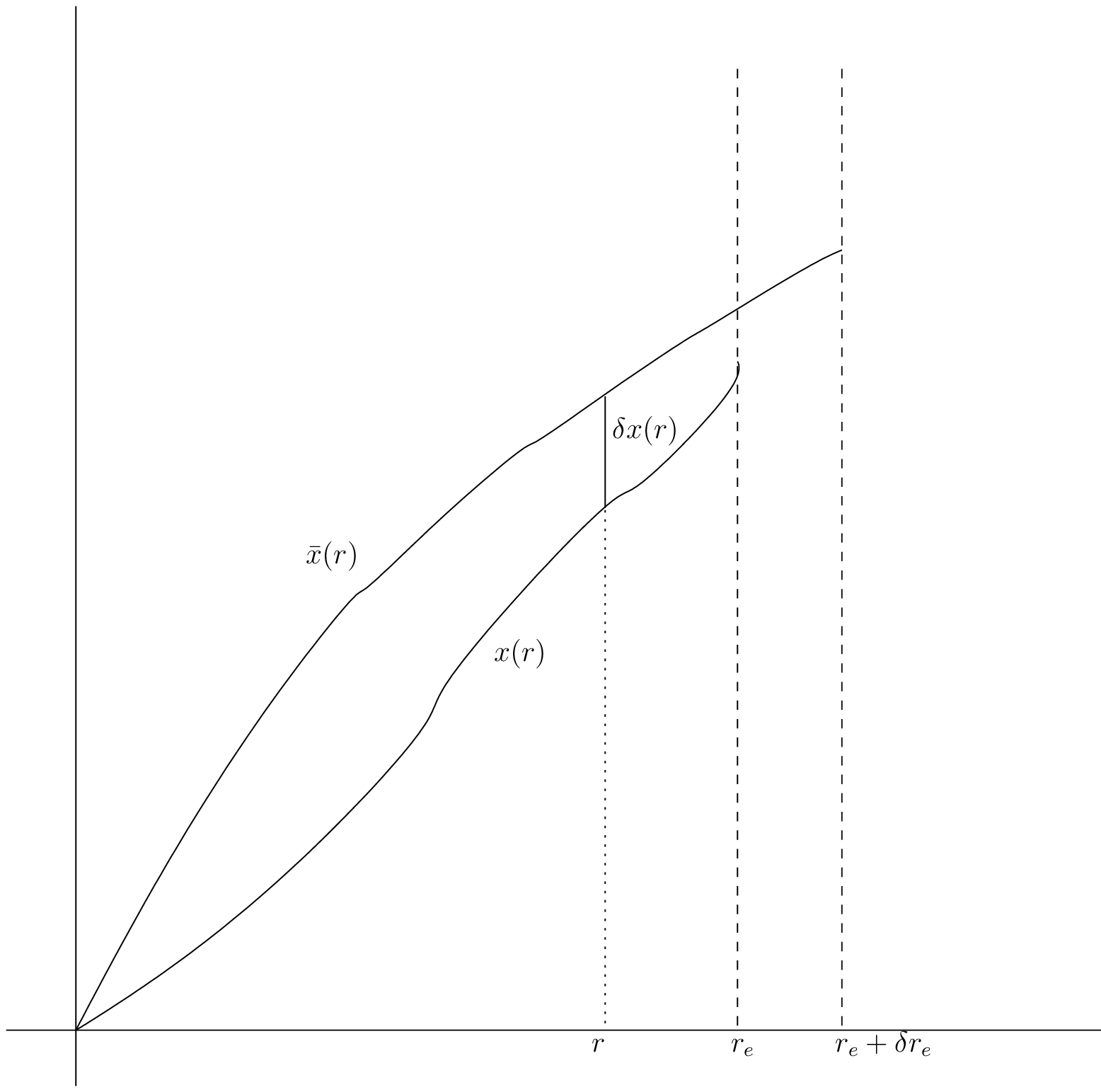}
\caption{The classical solutions corresponding to different boundaries.
}
\label{curve}
\end{center}
\end{figure}

Now we turn to variations in $r_e$. The idea here is that we vary the
boundary $r_e$, an look for variations in $x$ which result from
considering the extrema of the action with this new boundary. In other
words, in Figure \ref{curve}, both $x$ and $\bar x$ are solutions of the
Euler-Lagrange equations with their respective boundaries.

One way to handle this shift in boundary is to think of $r$ as a map
from $u\in [0,1]$ to the integration range (i.e., write $r$ as $r(u)$).
Then we can think of $u$ as the time variable, with fixed boundaries $0$
and $1$, and $r$ will be just another coordinate.
The changes in the boundary can be re-interpreted now as
variations in $r$, {\em at} the fixed boundaries of the $u$-interval.
Thus we have translated a moving boundary to a boundary term: something
easily handled using standard variational approaches (See e.g.,
chapter 1 of \cite{reuter} for a clean discussion of variational methods with boundary terms.). This
method\footnote{Incidentally, this
approach seems like a pretty general way to do variational mechanics.} works,
but unfortunately our action involves different pieces which contain
different
ranges for $r$: the upward strings go from $r_e$ to $\infty$, while the
downward strings are from $r_0$ to $r_e$. This makes this approach
somewhat complicated, so we will follow another path which is much
more direct and intuitive.

Let us consider what happens to the classical solutions of a system with
the action
\beq
S_0=\int_0^{r_e} L\left(x(r),x'(r)\right) dr,
\eeq
under variations of the boundary.
We are interested in finding new equations of ``motion" by
setting
\begin{eqnarray}
0=\delta_{r_e} S_0 = \int_0^{r_e+\delta r_e}L\left(\bar
x,\bar x'\right) dr-\int_0^{r_e} L\left(x,x'\right) dr
\end{eqnarray}
The important
thing to note here is that $\bar x$ is supposed to be
the classical solution corresponding to variations with the boundary fixed
at $r_e+\delta r_e$, just as $x$ is the classical solution with boundary
fixed at $r_e$. We have denoted $\bar x(r)-x(r) \equiv \delta x(r)$. Upto
first order in small quantities, we can write,
\begin{eqnarray}
0= \int_0^{r_e}\big(L(\bar x,\bar x')-L(x,x')\big)
dr+L(x,x')|_{r_e}\delta r_e.
\end{eqnarray}
By the usual tricks, the first piece can be massaged into the form
\begin{eqnarray}
\int_0^{r_e}dr\left(\frac{\partial L}{\partial x}\ \delta x+\frac{\partial
L}{\partial x'}\ \delta x'\right)=\int_0^{r_e} dr
\biggl[\left(\frac{\partial L}{\partial x} -
\frac{d}{dr}\frac{\partial L}{\partial x'}\right)\delta
x+\frac{d}{dr}\left(\frac{\partial L}{\partial x'} \ \delta
x\right)\biggr]. \nonumber
\end{eqnarray}
The Euler-Legrange term vanishes because $x$ is a classical path, and the other piece gets integrated and
receives contributions only from the boundary. Since we imagine that the
boundary at zero is held fixed, we end up with
\beq
0=\frac{\partial L}{\partial x'}\ \delta
x\bigg|_{r_e}+L(x,x')|_{r_e}\delta r_e.
\eeq
An important input at this stage comes from the fact that $\delta x(r)$
is the variation at fixed $r$ between classical solutions with
different boundaries. Notice that by definition $\delta_T x(r_e)=\delta
x|_{r_e} +
\frac{\partial x}{\partial r}\delta r\big|_{r_e}$ is zero: at the
new boundary $r_e+\delta r_e$, the new classical solution ${\bar x}$ is supposed to
have vanishing variations as well.
So we get the final result,
\beq
0=\left(L-\frac{\partial L}{\partial x'}x'\right)\bigg|_{r_e}.
\eeq
This was done for the case when the moving
boundary was at the upper end
of integration. For quarks hanging from the boundary of AdS, there will be
an overall sign. Putting all these ingredients together, we finally get
\begin{eqnarray}
-\sum_{m\in \{{\rm up}\}}\bigg(  {\cal L}_m-\frac{\partial{\cal
L}_m}{\partial x_{i,m}'} x_{i,m}'\bigg)\bigg|_{r=r_e} + \sum_{n\in \{{\rm
down}\}} \bigg({\cal L}_n-\frac{\partial{\cal
L}_n}{\partial x_{i,n}'}
x_{i,n}'\bigg)\bigg|_{r=r_e}+\frac{dS_{D5}}{dr_e}=0\hspace{0.2in}
\end{eqnarray}
as the $r_e$-equation of motion for the various quarks and the D5-brane.

\subsection*{{\bf B.} \ Trailing String}
\addcontentsline{toc}{subsection}{{\bf B} \ Trailing String}
\renewcommand{\theequation}{B.\arabic{equation}}

In order to get a handle on the various dissociated baryonic configurations, we need to understand the trailing string solution that extends from the D5-brane to the black hole horizon. We will show in this appendix that the most general string of this form in the $x_1-x_3-r$ space lies along the wind direction ($x_3-r$ plane). This intuitively natural conclusion is important because (due to the force balance conditions at the D5 brane) it considerably restricts the resultant dissociation configurations that are allowed.

The action for the most general static string stretched between the D5 and the horizon is,
\bea
S=\frac{{\cal T}}{2 \pi \alpha'}\int_{r_0}^{r_e}dr \sqrt{A\Big(\frac{1}{f}+\frac{r^2 x_1'^2}{R^2}\Big)+\frac{r^2f(r)x_3'^2}{R^2}}.
\eea
The equations of motion for the two components are
\bea
x_1'=\frac{R^2}{r}\frac{K_1}{\sqrt{r^2fA-R^2K_3^2A-R^2K_1^2f}}, \\
x_3'=\frac{R^2}{fr}\frac{K_3 A}{\sqrt{r^2fA-R^2K_3^2A-R^2K_1^2f}}.
\eea
By defining
\bea
t=\frac{r}{r_0}, \ \ \mu=\frac{\alpha}{\rho^2}, \ \ \nu=\frac{\beta}{\rho^2} \ \ {\rm and} \ \ x_{1,3}=\frac{R^2}{r_0}z_{1,3},
\eea
we can rewrite the above equations as
\bea
\left(\frac{dz_1}{dt}\right)^2&=&\mu^2\frac{1}{(t^4-\cosh^2 \eta)(t^4-1-\nu^2)-\mu^2(t^4-1)}, \\
\left(\frac{dz_3}{dt}\right)^2&=&\frac{\nu^2 \ (t^4-\cosh^2 \eta)^2}{(t^4-1)^2}\frac{1}{(t^4-\cosh^2 \eta)(t^4-1-\nu^2)-\mu^2(t^4-1)}.
\eea
We would like to have simultaneous solutions of these two equations. A real solution can clearly only exist if
\bea
(t^4-\cosh^2 \eta)(t^4-1-\nu^2)-\mu^2(t^4-1)>0.\label{meao}
\eea
(We will be sloppy about distinguishing $>$ and $\ge$ in what follows, we will consider the boundaries explicitly.)
The variable $t$ ranges from $t=1$ at $r=r_0$ (horizon) to $t=1/\rho$ at $r=r_e$ (D5 brane). We are interested in the case where $1/\rho > \sqrt{\cosh \eta}$ (See, for example, the screening length plots.). So for the entire range $1<t<\sqrt{\cosh \eta}+\epsilon$, with $\epsilon$ small enough\footnote{The integration limits are between 1 and $\frac{1}{\rho}$. We know that $\frac{1}{\rho}$ is bigger than $\sqrt{\cosh \eta}$, even though we don't know how much bigger. So at least for sufficiently small positive $\epsilon$, we can claim that the inequality should hold.} and positive, the above inequality should hold for the solution to make sense. Doing a variable redefinition, this means that
\bea
x^2+x(\sinh^2 \eta-\mu^2-\nu^2)-\mu^2 \sinh^2 \eta >0,
\eea
should hold for $ -\sinh^2 \eta < x < \delta $ with small, positive $\delta$. The above inequality is satisfied as long as $x$ is either less or more than {\em both} the roots of the quadratic\footnote{As opposed to being lesser than the bigger root and bigger than the smaller root.} (the discriminant is positive, so the roots are never complex). If one of these ranges overlaps with $ -\sinh^2 \eta < x < \delta $, then we have a solution. This happens iff
\bea
-\sinh^2 \eta > \frac{-(\sinh^2 \eta-\mu^2-\nu^2)+\sqrt{(\sinh^2 \eta-\mu^2-\nu^2)^2+4\mu^2\sinh^2 \eta}}{2},
\eea
or
\bea
\delta < \frac{-(\sinh^2 \eta-\mu^2-\nu^2)-\sqrt{(\sinh^2 \eta-\mu^2-\nu^2)^2+4\mu^2\sinh^2 \eta}}{2}.\label{second}
\eea
The former inequality implies
\bea
\nu^2 \sinh^2 \eta < 0.
\eea
This is clearly impossible, but there remains the possibility of the limiting case $\nu=0$, when the inequality is saturated. But this case is clearly pathological because it is easy to see that the original inequality (\ref{meao}) is violated for values of $t$ that lie close enough to 1. The remaining possibility is the second inequality (\ref{second}). This can be rewritten as
\bea
\delta^2+\delta (\sinh^2 \eta-\mu^2-\nu^2) - \mu^2 \sinh^2 \eta > 0,
\eea
and it needs to be satisfied if we tune $\delta$ small enough. This means that the only possibility is the
limiting case $\mu=0$, where the inequality is saturated. But this is precisely the case when the trailing string is purely along $x_3$, which gives rise to the solution considered in appendix A of \cite{Liu:2006he}.

\subsection*{{\bf C.} \ Other Configurations}
\addcontentsline{toc}{subsection}{{\bf C} \ Other Configurations}
\renewcommand{\theequation}{C.\arabic{equation}}

In this appendix we will look at some dissociated configurations to illustrate the fact that the energy plots can significantly differ from the ones considered before, depending on the configuration of quarks. The configurations we consider here are the minimal configurations allowed, if one stipulates that after a quark has dissociated, the remaining quarks in the baryon still remain on the circle. Of course, since we do not understand the dynamics, this is an {\em ad hoc} assumption. The purpose of this section is to give a flavor of the various dissociations patterns allowed.

\begin{figure}[h!]
\begin{center}
\includegraphics[width=0.5\textwidth
]{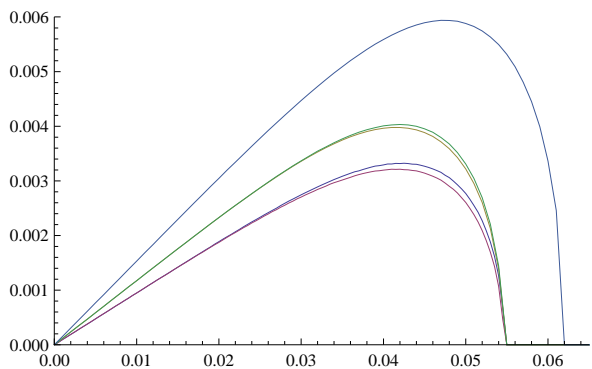}
\caption{$LT$ vs. $\rho$ for the various cases. The top curve is for the undissociated baryon. The color code for the rest is as follows. Violet: case B, blue: case A, green: case I (from section 3), brown:  case II. All the plots in this appendix are for $\eta=6$.
}
\label{x}
\end{center}
\end{figure}

\begin{figure}[h!]
\begin{center}
\includegraphics[width=0.5\textwidth
]{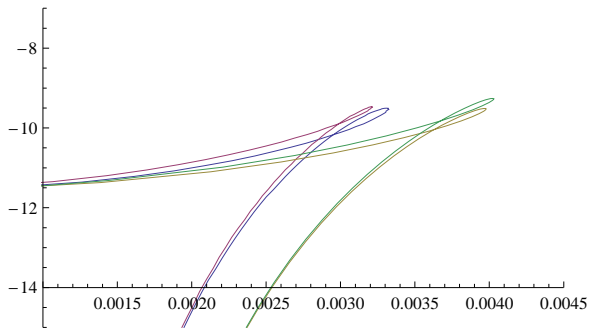}
\caption{$\frac{E}{T\sqrt{\lambda}}$ vs. $L$ at $\eta=6$. The color code is the same as before.
}
\label{y}
\end{center}
\end{figure}

\begin{figure}[h!]
\begin{center}
\includegraphics[width=0.5\textwidth
]{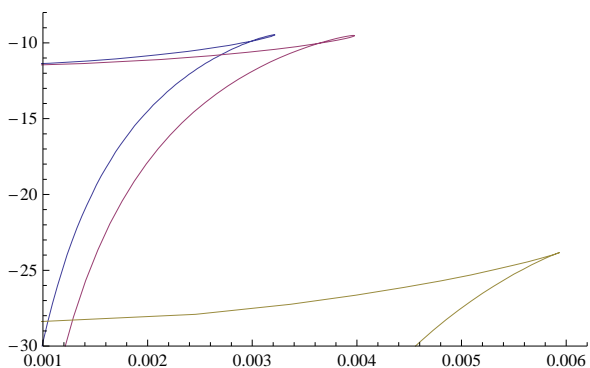}
\caption{$\frac{E}{T\sqrt{\lambda}}$ vs. $L$ at $\eta=6$. The lowest curve is undissociated baryon, the rest of the color code is the same as before.
}
\label{z}
\end{center}
\end{figure}

The configurations we consider are easily described by the condition that the quarks remain on the circle. The force balance conditions for our simple system are stringent enough that if the dissociated quark is one of the transverse ones, then not all of the remaining three quarks can remain in a circle. So we will look at the two cases where one of the quarks in the longitudinal direction is the dissociated one\footnote{There are a few configurations one can consider in order to get an idea about transverse quark dissociations. One of them was already considered in section 3. Many other possibilities are ruled out or become uninteresting because of the constraints on the configuration. For example, the case where all three undissociated quarks are at the same location at the boundary, is fully fixed by force balance and symmetry. Also, it does not have the analog of a screening length, so it is not very interesting for our purposes.}. In the $(x_1,x_3)$ plane, the two cases we consider will be defined by the quark positions
\bea
{\rm Case \ A:} && (1,0), (-1,0), (0,1), \\
{\rm Case \ B:} && (1,0), (-1,0), (0,-1).
\eea
For example, in this notation, the longitudinal case (case I) considered in the main text will be defined by the quark locations $ (0,1),(0,0),(0,-1)$.

We will not present the details of the computations because the general formalism is the same as in the examples we considered in the main body of the paper. The one slight subtlety is that to make sure that the boundary quarks are arranged a circle, the relations we need to impose are,
\bea
L_{x_3}^{(2)}-L_{x_3}^{(1 \ {\rm or} \ 3)} = L_{x_1}^{(1 \ {\rm or} \ 3)}
\eea
for case A, and
\bea
L_{x_3}^{(1 \ {\rm or} \ 3)}-L_{x_3}^{(2)} = L_{x_1}^{(1 \ {\rm or} \ 3)},
\eea
for case B. The $L_{x_i}$ are measured from the D5-brane, and the superscripts denote the relevant quark (notice the $1 \leftrightarrow 3$ symmetry).
We present here the plots of the energies, screening lengths, together with those of the cases considered in section 3. The comparisons with the undissociated baryons are also presented.

\newpage

\end{document}